\documentstyle [preprint,eqsecnum,aps]{revtex}
\tightenlines
\begin{document}
\draft
\title{Brillouin scattering studies in Fe$_3$O$_4$ across the Verwey transition}
\author{Md. Motin Seikh$^{1}$, Chandrabhas Narayana$^{2}$, P.A. Metcalf$^3$, J.M. Honig$^3$ and A. K. Sood$^{4}$} 
\address{$^1$ Solid State Structural Chemistry Unit, Indian Institute of Science, Bangalore 560 012, India. \\
$^2$ Chemistry and Physics of Materials Unit, Jawaharlal Nehru Center for Advanced Scientific Research, Jakkur P.O., Bangalore 560 064, India.\\
 $^3$Department of Chemistry, Purdue Univesity, West Lafayette, Indiana 47907, USA.\\
$^4$ Department of Physics, Indian Institute of Science, Bangalore 560 012, India. \\}

\date{\today}
 
\maketitle

\begin{abstract}
Brillouin scattering studies have been carried out on high quality single crystals of Fe$_3$O$_4$ with [100] and [110] faces in the temperature range of 300 to 30 K.  The room temperature spectrum shows a surface Rayleigh wave (SRW)  mode at 8 GHz and a longitudinal acoustic (LA) mode at 60 GHz. The  SRW mode frequency shows a minimum at the Verwey transition temperature $T_V$ of 123 K.  The softening of the SRW mode frequency from about 250 K to $T_V$ can be quantitatively understood as a result of a decrease in the shear elastic constant C$_{44}$, arising from the coupling of shear strain to charge fluctuations.  On the other hand, the LA mode frequency does not show any significant change around $T_V$, but shows a large change in its intensity.  The latter shows a maximum at around 120 K in the cooling run and at 165 K in the heating run, exhibiting a large hysteresis of 45 K.  This significant change in intensity may be related to the presence of stress-induced ordering of Fe$^{3+}$ and Fe$^{2+}$ at the octahedral sites, as well as to stress-induced domain wall motion.
\end{abstract}


\narrowtext

\section{Introduction}
Magnetite has attracted renewed attention in recent years because of its interesting electronic, magnetic, and transport properties as well as potential applications in the magnetic multilayer devices \cite{shepherd1985,shepherd1991,aragon1988,anderson1997}.  Stoichiometric Fe$_3$O$_4$ shows a first order metal-insulator (MI) transition, referred to as Verwey transition, with the reduction in the electrical conductivity of nearly two orders of magnitude at the transition temperature ($T_V$).  The cubic inverse spinel structure (F3dm) of Fe$_3$O$_4$ may be written as Fe$^{3+}$[Fe$^{2+}$Fe$^{3+}$]O$_4$, where the tetrahedrally coordinated Fe$^{3+}$ ions (A-site) is differentiated from the bracketed Fe$^{2+}$ and Fe$^{3+}$ octahedral sites (B-sites). The spins of the two B-site cations are antiparallel to the single A-site Fe$^{3+}$ cation, giving rise to ferrimagnetic order, with a transition temperature of 850 K \cite{pauthenet1950}. Cubic magnetite at room temperature changes to a monoclinic structure below $T_V$, containing four rhombohedrally distorted cubic cells  \cite{yoshida1979,zuo1990}. However, there is also an evidence of a triclinic structure at low temperatures \cite{miyamoto1988,medrano1999}.  The question of  charge ordering (CO) of the Fe$^{2+}$ and Fe$^{3+}$ states on the B sites in the low temperature phase is a matter of continued controversy.  The recent experimental data on M\"{o}ssbauer \cite{berry1998}, neutron \cite{iizumi1975,iizumi1982}, x-ray diffraction \cite{yoshida1979} and $^{57}$Fe NMR \cite{novak2000} studies are not consistent with the expected CO.  On the other hand, the high-energy transmission electron diffraction studies show charge ordering below the Verwey transition \cite{zuo1990}.  Also, the high resolution x-ray and neutron powder diffraction data has revealed charge ordering, two Fe ions with a charge of +2.4 and the other two of  +2.6,  as evidenced by the presence of two types of BO$_6$ octahedra and the distribution of B-B distances \cite{wright2001,wright2002}.

The effect of nonstoichiometry, either by oxygen deficiency or impurities has been studied recently \cite{honig1995,kozlowski1996,kozlowski2000,kozlowski1997,kakol1989,kakol1991,aragon1992}. Nonstoichiometry shifts the Verwey transition to lower temperatures as well as broadens the transition \cite{honig1995}.  The nature of the transition also changes from first order to higher order beyond a certain doping level \cite{shepherd1991}.  The influence of the linear magnetoelastic interaction on the ultrasonic sound velocity and attenuation in Fe$_3$O$_4$ was investigated by Moran and L\"{u}thi \cite{moran1969}.  Of the three elastic constants, only C$_{44}$ showed a significant softening from room temperature to $T_V$, whereas C$_{11}$ and C$_{12}$ showed an increase as $T$ was lowered, with a jump at $T_V$ \cite{schwenk2000}.

The role of lattice in the Verwey transition in Fe$_3$O$_4$ has been discussed in the literature \cite{iizumi1982}.  Recent Raman studies have revealed a strong electron phonon coupling for the T$_{2g}$ optical phonon \cite{gupta2002}.  The motivation of the present work is to study the behavior of acoustic phonons across the Verwey transition using Brillouin scattering experiments.   In this paper, we present the temperature dependence of the surface as well as bulk acoustic phonons in pristine Fe$_3$O$_4$ from 300 to 30 K.  A pronounced softening of the surface acoustic phonon velocity as $T$ is lowered to $T_V$, and its subsequent significant hardening below the transition, is observed in our experiments. On the other hand, there is no appreciable change in the bulk acoustic phonon velocity with temperature along both  [100] and [110] directions.  However, a large change in the intensity for the bulk mode is observed near $T_V$, with a hysteresis of about 45 K.  This non-monotonic behavior of the intensity is may be due to the large domain wall movements expected close to $T_V$ as well as stress induced ordering of the ions. 

\section{Experimental Details}
Single crystals of magnetite used in this study were grown by the standard skull melting technique \cite{harrison1984,wittenauer1995}.  Standard four probe resistivity measurements show the MI Verwey transition to occur at 123 K \cite{guptathesis}. The crystals were mechanically polished using diamond suspension to obtain a highly reflecting surface. After polishing, the crystals were cleaned in methanol, followed by acetone, using an ultrasonic bath. The measurements have been performed on the [100] and [110] faced crystals of the undoped Fe$_3$O$_4$.  Brillouin spectra were recorded in the oblique-incidence as depicted \cite{sandercock1982,stageman1984} 180$^\circ$-back-scattering geometry, with an incident angle of $\theta = 45^\circ$ relative to the surface normal.  Ideally $\theta$ should be $70^{\circ}$ for an oblique-incidence experiment. In our experiment, the horizontal plane defines the scattering plane and the sample is mounted with its surface perpendicular to the scattering plane. The optical windows in our cryostat used for the temperature dependent studies are placed at $90^\circ$ to each other. Thus the angle of incidence has to be $45^{\circ}$, in order to avoid hitting the reflected light inside the cryostat. The scattered light was analyzed using a JRS Scientific instruments (3+3)-pass tandem Fabry-Perot interferometer equipped with a photo avalanche diode detector. The 532 nm ($\lambda$) line of Nd:YAG single mode solid-state diode-pump frequency doubled laser (Model DPSS 532-400, Coherent Inc. USA) was used as the excitation source. The temperature-dependent measurements with a temperature accuracy of $\pm 1$ K were carried out inside a closed cycle helium cryostat (CTI Cryogenics, USA).  The incident laser power was kept at $\sim$ 25 mW focused to a spot of diameter $\sim$ 30 $\mu$m. The typical time required per spectrum is 1.5 h. The lineshape parameters - peak position, full-width at half maximum (FWHM), and area were extracted by non-linear least square fitting of the data with Lorentzian functions, together with appropriate background corrections.

\section{Results and Discussion}
Figure 1 shows the room temperature Brillouin spectrum of Fe$_3$O$_4$ [100], displaying two well resolved peaks, at 8 GHz and 60 GHz. The 8 GHz mode shows a linear dependence of its frequency on the wavevector component parallel to the surfaces, $q_\parallel$ ($= 4 \pi $sin$\theta / \lambda$, where $\theta$ is the scattering angle), expected for a surface Rayleigh wave (SRW) (as shown in inset (a) of Fig. 1).  Further, the frequency of the 8 GHz mode shows a sinosidal dependence (see inset (b) of Fig. 1) on the azimuthal angle $\phi$ (angle of rotation with respect to the scattering plane about the normal), confirming it to be a SRW mode. For the [100] oriented crystal, the high frequency mode appears in HH polarization, showing it to be a bulk longitudinal acoustic (LA) mode.  Here H stands for horizontal polarization, which lies in the scattering plane. The surface Rayleigh wave velocity (obtained from the slope of the curve shown in inset (a) of Fig. 1) is 3200 m/sec, and the bulk longitudinal acoustic velocity is 6600 m/sec, in close agreement with the earlier reported longitudinal sound velocity along [100] direction \cite{moran1969}.  We point that a very weak mode is seen at ~ 13 GHz in unpolarized spectrum, which can be attributed to the magnon.  This mode could not be observed with the sample inside the cryostat.  We further note that transverse acoustic phonons, though seen in oblique incidence Brillouin scattering experiments \cite{stageman1984,sandercock1978} are too week to be observed in our backscattering experiments.  We believe that the strong absorption at 2.32 eV excitation energy \cite{balberg1971} broadens the Brillouin line making it difficult to observe.

The temperature-dependence of the SRW frequency, $\nu_s$, along [100] direction of Fe$_3$O$_4$ is shown in Fig. 2(a) .  Upon cooling below 250 K, there is a large softening in $\nu_s$ ($\sim$ 6 \%) till $T_V$.  Below $T_V$ there is a steep increase in $\nu_s$ ($\sim$ 10 \%).   The temperature-dependence of SRW along [110] direction is similar to that observed in [100] direction as shown in Fig. 2(b), though the magnitude of softening is smaller compared to the [100] crystal.   The magnitude of anisotropy in $\nu_s$ in the two directions [100] and [110] is shown as a function of temperature in the inset of Fig. 2(b).  The full width at half maxima (FWHM) as well as intensity of the SRW mode for both directions [100] and [110] show changes near $T_V$. The FWHM of the surface mode is comparable to that of the instrumental broadening and hence cannot be qualitatively interpreted.  However, we note that below $T_V$, the full width at half maxima of the SRW increases, which is still to be understood.    The intensity of the SRW mode decreases as the temperature is lowered due to the Bose-Einstein population factor.  The SRW mode could not be observed below 50 K in the cooling cycle, but it reappears upon heating the sample above 50 K.  

The bulk LA velocity does not undergo any observable change  with temperature, {\it i.e.}, there is no appreciable change in the Brillouin peak position (as shown in Fig. 3(a)) or FWHM. However, the peak intensity shows a strong anomaly around $T_V$ for both the [100] as well as the [110] directions, as can be seen in Fig. 3(b) (for the [100] direction).    In the cooling run, the intensity gradually increases with a decrease in temperature and shows a maximum at $\sim$ 120 K, followed by a sharp drop below $T_V$ (see filled circles in Fig. 3(b)); the mode cannot be observed below 70 K.  In the heating run, the LA mode reappears at $\sim$ 70 K and its intensity is a maximum at 165 K (see open circles in Fig. 3(b)), displaying a considerable hysteresis of  $ \sim$ 45 K (Fig. 3(b)).  The hysteresis is approximately double that which is measured in the magnetization measurements \cite{domenicali1950}.  We now explain our observations.

\subsection{Surface Rayleigh wave}
For the cubic crystal, the SRW frequency depends on the elastic constants C$_{11}$, C$_{12}$ and C$_{44}$ \cite{cummins1971,vacher1972}.  Using Green's function method \cite{comins01,zhang98} we have calculated surface mode spectra for different temperatures in the [100] direction with the elastic constants C$_{11}$, C$_{12}$ and C$_{44}$  taken from Ref. \onlinecite{schwenk2000}, and shown by dotted lines in the inset of Fig. 2(a).  The values of $\nu_s$ thus calculated are plotted as dotted line in Fig. 2(a).   Though the temperature dependence of the calculated $\nu_s$ is similar to our observation, the quantitative agreement is not so good.  It is important to note that the values of the elastic constants shown in the inset of Fig. 2(a) were obtained in a magnetic field of 0.5 T.  In order to obtain a better fit of the calculated $\nu_s$ with the data, we adjusted the value of C$_{44}$, keeping the values of C$_{11}$ and C$_{12}$ as before.  This was done because the temperature dependence  of $\nu_s$ is similar to that of C$_{44}$.  The solid line in Fig. 2(a)  shows the calculated $\nu_s$, in very good agreement with the data.  The value of C$_{44}$ needed to get this good fit are shown by filled circles in the inset of Fig. 2(a).

We will now address the softening of C$_{44}$ between 250 K and $T_V$.  This behavior of C$_{44}$ has been attributed to a bilinear coupling of the shear strain, $\epsilon$, with the order parameter, $\eta$, related to the charge ordering processes \cite{moran1969,schwenk2000}.  A similar behavior of the elastic constant C$_{44}$ has been observed in other charge ordered systems Yb$_4$As$_3$ \cite{goto1999} and ${\alpha^\prime}$-NaV$_2$O$_5$ \cite{schwenk1999}.  Taking the usual Landau free energy expansion in terms of an order parameter $\eta$, along with the additional strain-order parameter coupling term $F_C$ leads to the free energy expression $F = F_0+\alpha\eta^2 + \beta\eta^4 + F_C + F_{el}$.  Here $F_0$ is the part of the free energy which remains constant through the phase transition, $F_{el}$ is due to the elastic energy, and $F_C = g \epsilon \eta$, where $g$ is the coupling constant between the strain and the charge fluctuation order parameter. This coupling leads to the softening of the shear elastic constant \cite{goto1999} 
\begin{equation}
C_{44}(T) = C^0_{44}(T-T_0)/(T-\Theta)
\label{eq1}
\end{equation}
where $\Theta$ is the transition temperature without the strain-order parameter coupling, and
\begin{equation}
T_0 = \Theta + 0.5 g^2  \alpha^\prime C_{44}
\label{eq2}
\end{equation}
with the usual Landau term $\alpha = \alpha^\prime (T - \Theta)$. The difference of the two characteristic temperatures $(T_0 - \Theta)$ corresponds to the coupling strength of the strain and the order parameter \cite{goto1999}.  The values of C$_{44}$,  obtained by the fit of the SRW frequency  at various temperatures shown by the filled circle in the inset of Fig. 2(a) were fitted to give $T_0$ = 57 K and $\Theta$ = 40 K, showing the coupling strength between strain and order parameter to be around 17 K, which is somewhat higher than that estimated ($\sim$ 10 K) by Schwenk et al \cite{schwenk2000}.  Since the contribution of C$_{44}$ to $\nu_s$ in [100] and [110] directions are different, the anisotropy parameter plotted in the inset in Fig. 2(b), will also reflect the temperature dependence of C$_{44}$.  The gradual decrease in intensity with temperature is expected from the usual thermal population factor.

\subsection{Bulk LA mode}

Unlike transparent materials, absorbing materials have a complex refractive index, $n = \eta + i \kappa$ and the optical field falls off exponetially in the medium.  This results in the uncertainity in the wavevector transfer $q$ \cite{sandercock1972,loudon1978,dresselhaus1975}, leading to broadening of the Brillouin lines, over and above the lifetime broadening.  The full-width at half maximum (FWHM), $\Gamma = 4 v k_\circ \kappa$ and the peak position at $\omega = 2 \eta v k_\circ$ gives $\frac{\Gamma}{\omega} = 2 \frac{\kappa}{\eta}$;  where $v$ is the sound velocity, $k_\circ$ is the incident wavevector \cite{sandercock1982,boukari1998}.  For Fe$_3$O$_4$, $\alpha$ has been measured only up to 0.75 eV and its value is $\sim 3.5 \times 10^{3}$ cm$^{-1}$ \cite{balberg1971}.   Unfortunately, there is no measurement of $\alpha$ at 2.32 eV.  Extrapolating the results of Balberg et al \cite{balberg1971}, we find $\alpha \sim 10^{5}$ cm$^{-1}$ at 2.32 eV.  This gives $\Gamma = 14$ GHz, which is close to the observed value of 16 GHz in our experiments.  Hence, most of the width of the bulk LA mode is due to the large absorption coefficient of Fe$_3$O$_4$. 

Schwenk et al. \cite{schwenk2000} have shown that the change in C$_{11}$ is $\sim$3 \% between 300 to 80 K (see inset in Fig. 2(a)), which implies a change of  1.5 \% ($\sim$ 0.9 GHz) in LA frequency, since the change in density is negligible \cite{elastic}. This is within the resolution limit of 1 GHz  in our Brillouin experiments and is consistent with the temperature dependence shown in Fig.  3(a).  Now we give a plausible explaination of the non-monotonic variation of intensity of the LA mode with temperature (Fig. 3(b)).  The intensity of the scattered light by the LA mode is related to the Pockels coefficients $p_{11}$ and $p_{12}$, which are related to the derivatives of the dielectric function with respect to the strain $\epsilon$.  In the case of Fe$_3$O$_4$, the strains are not completely elastic.  There are two sources of non-elastic strains; one is a stress-induced domain wall motion $\epsilon_d$ (magnetostriction) \cite{bozorth1951,becker1939}, and the other is stress-induced ordering of Fe$^{2+}$ and Fe$^{3+}$ among the octahedral sites, $\epsilon_{order}$.  Hence the total strain $\epsilon$ is the sum of  $\epsilon_{elastic}$,  $\epsilon_{d}$ and $\epsilon_{order}$.  Fine et al. \cite{fine1954} have found that the Young's modulus $Y$ ($Y$ = stress/$\epsilon$) of magnetite shows a minimum near $T_V$ (see inset of Fig. 3(b)), which is absent in a magnetic field.  Since a magnetic field should have no effect on elastic strains and should not prevent ordering, the anomaly in the Young's modulus is attributed to the contribution of the domain wall motion to the strain.   Bickford et al. \cite{bickford1953} suggested that if a single crystal of Fe$_3$O$_4$ is cooled through the transition in absence of magnetic field, it is not a true single crystal below $T_V$ . Furthermore, the regions may be twinned, since the a and  b crystal axes can be interchanged.   In some of the earlier experiments the domain size has been found to be around 0.5 up to 2 $\mu$m \cite{ziese2002,kalev2003}.  This is much smaller compared to the laser spot of 30 $\mu$m diameter. However, the crystal reverts to the single crystalline cubic phase upon heating above $T_V$ \cite{bickford1953}.   An increase in strains around $T_V$ is expected to result in an increase in the value of the Pockel coefficients and hence, the intensity of the LA mode.  It would be interesting to measure the intensity of the mode in the presence of magnetic field, which will make the sample single domain. 

In conclusion, Brillouin scattering studies on Fe$_3$O$_4$ reveal that there is a large change in the surface Rayleigh wave velocity associated with the Verwey transition, which is driven by a bilinear coupling of shear strains with the charge ordering.  The transition temperatures T$_0$ and $\Theta$ are found to be 57 K and 40 K, showing the coupling strength to be 17 K.    The temperature behavior of the intensities of the longitudinal acoustic mode may be understood in terms of non-elastic strains due to domain wall motion and charge ordering. 

Mr. Seikh would like to thank Council of Scientific and Industrial Research (CSIR), India for a research fellowship. We like to acknowledge the continued support, encouragement and fruitful discussions with Prof. C.N.R. Rao.  Thanks are due to Department of Science and Technology (India) for financial support.

\begin{figure}
\caption{Room temperature Brillouin spectrum of Fe$_3$O$_4$ along [100] direction showing both the surface Rayleigh wave (SRW) at 8 GHz and bulk mode (B) at 55 GHz. Inset (a) and (b) shows the wave vector dependence and angular variation of SRW mode velocity, respectively.}
\label{fig1} 
\end{figure}

\begin{figure}
\caption{Temperature dependence of the SRW mode frequency (a) along [100] direction  and (b) along [110] direction for Fe$_3$O$_4$.  The dotted line in (a) is the calculated Brillouin shifts using the Green's function, taking C$_{11}$, C$_{12}$  and C$_{44}$ from Ref. [24] shown by dotted lines in the inset.  The solid line in (a) is calculated as a best fit to the data in Ref. [24], but varying C$_{44}$ (as shown by the filled circle in the inset).  The solid line in the inset is a fit to C$_{44}$ (filled circle) using Eq. (1).  Inset in (b) shows the anisotropy  in frequency along [100] and [110] directions as a function of temperature.} 
\label{fig2} 
\end{figure}

\begin{figure}
\caption{Temperature dependence of the bulk LA mode (a) frequency and (b) intensity of Fe$_3$O$_4$ along the [100] direction.  The solid (open) circles represent cooling (heating) run data.  The solid lines drawn in (b) are a guide to the eye.  The inset in (b) shows the Young's modulus along [100], reproduced from Ref. [24].}
\label{fig3}   
\end{figure}

\begin{references}
\bibitem{elastic} $\Delta v/v = 1/2 \Delta C_{ij}/C_{ij} - 1/2 \Delta \rho/ \rho$, where $v$ is the sound velocity of the acoustic mode, C$_{ij}$ is the elastic constant associated with $v$ and $\rho$ is the density.
\end{references}
\end{document}